# PREDICTIVE NON-EQUILIBRIUM SOCIAL SCIENCE

Rich Colbaugh

Sandia National Laboratories
Albuquerque, NM USA
colbaugh@comcast.et

Kristin Glass

New Mexico Tech
Socorro, NM USA
kglass@icasa.nmt.edu

Curtis Johnson

Sandia National Laboratories
Albuquerque, NM USA
cjohnso@sandia.gov

**ABSTRACT**

Non-Equilibrium Social Science (NESS) emphasizes dynamical phenomena, for instance the way political movements emerge or competing organizations interact. This paper argues that predictive analysis is an essential element of NESS, occupying a central role in its scientific inquiry and representing a key activity of practitioners in domains such as economics, public policy, and national security. We begin by clarifying the distinction between models which are useful for prediction and the much more common explanatory models studied in the social sciences. We then investigate a challenging real-world predictive analysis case study, and find evidence that the poor performance of standard prediction methods does not indicate an absence of human predictability but instead reflects (1.) incorrect assumptions concerning the predictive utility of explanatory models, (2.) misunderstanding regarding which features of social dynamics actually possess predictive power, and (3.) practical difficulties exploiting predictive representations.

## 1 INTRODUCTION

Non-Equilibrium Social Science (NESS) emphasizes dynamical phenomena, for example the way collective behaviors emerge and evolve or how competing entities interact (NESS 2012). From a practical point of view, one of the most compelling reasons to study NESS is to learn enough to be able to form useful *predictions*. For instance, in domains such as economics, public policy, human health, and national security, analysts and advisors are often asked to forecast the eventual outcomes of social processes ranging from financial crises to revolutions. The task of inferring the existence and nature of activities which are already underway but are hidden in some way, sometimes referred to as "predicting the present" (Choi and Varian 2009), is also crucial in many applications (see, e.g., (Colbaugh and Glass 2012a) for a review of these two classes of prediction problems).

Despite its central importance to practitioners, prediction is only a peripheral concern in many social science disciplines. Finance provides an illustrative example, with market participants and analysts focusing almost exclusively on prediction while academic researchers concentrate mainly on explanation. One result of this dichotomy has been the development of explanatory models and theories which are considered pillars of research and yet afford very little predictive power or practical utility (Welch and Goyal 2008; Farmer, Patelli, and Zovko 2005). The situation is similar in other business-oriented fields. For instance, (Shmueli and Koppius 2011) shows that, despite substantial interest in prediction among practitioners, only 52 of the 1072 papers published in top-ranked management journals during the period 1999-2006 make predictive claims, and of these only seven "carried out proper predictive modeling or testing". Similar conditions exist in political science (Schrodt 2010), economics (Feelders 2002), sociology (Watts 2011), and elsewhere in social science (Berk 2008).

One possible explanation for this state of affairs could be that prediction, while undeniably important in practice, is somehow "unscientific" and therefore not worthy of serious attention by researchers (Kendall and Stuart 1977, Parzen 2001). However, careful and systematic examination reveals that predictive modeling and testing serve several essential scientific functions. At a fundamental level, philosophers of science have long argued that both explanatory and predictive power are required in a proper scientific theory, and indeed that explanation without predictive power is "pre-scientific" (Hempel and Oppenheim 1948). Concrete examples of the foundational role of prediction in the social sciences are given in (Hempel and Oppenheim 1948, Schrodt 2010; Ward, Greenhill, and Bakke 2010). Additional discussion concerning the value of prediction to the advancement of scientific theory may be found in (Shmueli and Koppius 2011) and the references therein.

Equally importantly, predictive methods enable the "operationalization" of key scientific tasks. For instance, the recent availability of large, rich datasets capturing myriad aspects of human behavior, such as the electronic traces of communication, innovation, consumption, and mobility left online, offer unprecedented views of human activity (Glass and Colbaugh 2011). Making sense of these data involves discovery and exploration of complex relationships and patterns which are difficult to hypothesize about *ex ante*, as is typically done with explanatory investigations, and is more naturally approached using predictive methods (Hastie, Tibshirani, and Friedman 2009). The development of new scientific theory requires rigorous empirical techniques for comparison and evaluation, for example of competing theories or of the discrepancies between a given theory and the real-world, and predictive models are well-suited to this type of analysis. Additionally, the concepts and methods of predictive analysis permit characterization of important aspects of NESS phenomena. Consider, as but one example, how knowledge of the *predictability* of dynamical processes has emerged as a fundamental element of scientific knowledge, allowing precise classification of the "know-ability" of attributes of systems and activities (Taleb 2007).

In view of the above discussion, this paper adopts the position that predictive analysis is an essential element of social science generally and NESS specifically, occupying a central role in scientific inquiry and representing a key activity of practitioners in domains such as economics, public policy, business, and national security. We begin, in Section 2, with a demonstration that predictive models and explanatory models are indeed different tools, possessing distinct objectives and implementing different strategies to manage inevitable modeling trade-offs. In particular, it is shown that good explanatory performance need not imply significant predictive power. Section 3 then presents a fairly comprehensive case study of an important class of social networks and their dynamics. The objects of interest are *signed* social networks, where positive and negative edges reflect friendly and antagonistic social ties, respectively. We derive a novel algorithm for edge-sign prediction that leverages structural balance theory (Heider 1948). The proposed algorithm outperforms a "gold-standard" method in empirical tests with two large-scale online social networks, with the boost in prediction accuracy being especially significant in situations where only limited training data are available. Interestingly, the inferred edge-signs are also shown to be useful when predicting the evolution of adversarial network dynamics. This case study offers evidence that the poor performance of traditional prediction techniques when applied to social phenomena does not indicate an absence of human predictability, but instead reflects misunderstanding regarding which social science models and features actually possess predictive power and how these models and features can be exploited in practical settings. Finally, in Section 4, we present a brief summary of the paper and suggest directions for future research.

## 2 PREDICTIVE V. EXPLANATORY MODELS

This section begins by identifying key differences in the objectives of predictive and explanatory analysis, then briefly examines the "bias-variance trade-off" of statistical modeling and its implications for prediction, and finally illustrates the importance of these trade-offs through a real-world example.

### 2.1 Analysis

In the social sciences, quantitative (e.g., statistical) models are used almost exclusively to realize explanatory goals, usually to test causal theories. In such models, a set of underlying factors which are measured by a vector of variables **X** are assumed to cause some effect of interest, quantified by the variable Y. The models are nearly always association-based, with regression models being the most common, and the approach is justified by asserting that the theory itself provides the causality. (This strategy is in contrast to one, for example, which employs a statistical method to infer causality more directly (Hastie, Tibshirani, and Friedman 2009)). Thus, in this paper, an *explanatory model* is one intended to test a causal explanation offered by social science theory. Alternatively, the goal of a *predictive model* is to predict new (e.g., future or hidden) events or activities, which in the present setup involves predicting the output Y corresponding to a newly observed set of input values **X**.

The importance of this difference in objectives lies in the well-known fact that measured observations do not provide a perfect representation of the underlying phenomena. Consider a social science theory postulating that variables $\mathbf{X} = [X_1, \ldots, X_n]^T$ cause variable Y via some relationship, and suppose that this relationship is "operationalized" in terms of a statistical model, say $E(Y) = f(\mathbf{X})$ (where $E(.)$ denotes expectation). In this setting, the objective of an explanatory model is to uncover the true underlying relationship $f(.)$, and the data (Y, **X**) are used to achieve that end, perhaps through regression. In contrast, in predictive modeling the focus is on the data (Y, **X**), and in particular on predicting the Y corresponding to a new observation **X**, and $f(.)$ is a tool for generating these predictions. Explanatory modeling and predictive modeling handle measurement noise differently in order to achieve these distinct goals.

As suggested above, good explanation requires an accurate estimate for $f(.)$ while it may be the case that good prediction is obtained with a different model, perhaps one that is less "accurate" but more parsimonious. To see how this can happen, let the phenomenon of interest be described by

$$Y = f(\mathbf{X}) + \varepsilon, \tag{1}$$

where $E(\varepsilon) = 0$, and denote by $f_{est}$ an estimate for $f(.)$, for instance obtained by fitting model $Y = f_{est}(\mathbf{X})$ to some "training" data $\{\mathbf{X}_i, Y_i\}_{i=1}^n$. Given a new observation **x**, our prediction for Y is computed as $f_{est}(\mathbf{x})$, and the expected mean square error (EMSE) associated with this prediction is:

$$\begin{aligned}
\text{EMSE} &= E(Y - f_{est}(\mathbf{x}))^2 \\
&= E(Y - f(\mathbf{x}) + f(\mathbf{x}) - f_{est}(\mathbf{x}))^2 \\
&= E(Y - f(\mathbf{x}))^2 + E(f(\mathbf{x}) - f_{est}(\mathbf{x}))^2 + 2E[(Y - f)(f - f_{est}(\mathbf{x}))] \\
&= E(\varepsilon^2) + E(f(\mathbf{x}) - f_{est}(\mathbf{x}))^2 \\
&= E(\varepsilon^2) + E[f(\mathbf{x}) - E(f_{est}(\mathbf{x})) + E(f_{est}(\mathbf{x})) - f_{est}(\mathbf{x})]^2 \\
&= E(\varepsilon^2) + E(f(\mathbf{x}) - E(f_{est}(\mathbf{x})))^2 + E(E(f_{est}(\mathbf{x})) - f_{est}(\mathbf{x}))^2 \\
&\quad + 2E[(f(\mathbf{x}) - E(f_{est}(\mathbf{x}))) (E(f_{est}(\mathbf{x})) - f_{est}(\mathbf{x}))] \\
&= E(\varepsilon^2) + E(f(\mathbf{x}) - E(f_{est}(\mathbf{x})))^2 + E(E(f_{est}(\mathbf{x})) - f_{est}(\mathbf{x}))^2. \tag{2}
\end{aligned}$$

Thus the EMSE consists of three components:

- $E(\varepsilon^2)$, the variance inherent in the phenomenon (1) – this is the prediction error which results even if the model is specified correctly and estimated perfectly;
- $E(f - E(f_{est}))^2$, the square of the model bias – this is the error arising from misspecification of the model;

- $E(E(f_{est}) - f_{est})^2$, the variance resulting from using a sample (the training data) to estimate $f_{est}(.)$.

The decomposition (2) quantifies the bias-variance trade-off – it is often the case that reducing model bias (second term) leads to an increase in estimation variance (third term) Hagerty and Srinivasan 1991).

In explanatory modeling, the emphasis is on minimizing model bias to obtain the most accurate representation for phenomenon (1), since this usually reflects some social theory of interest. Predictive modeling, on the other hand, seeks to minimize the combination of bias and estimation variance, which sometimes means employing misspecified models. Simple examples illustrating situations in which this occurs are discussed in (Hagerty and Srinivasan 1991; Hastie, Tibshirani, and Friedman 2009). In what follows we present a real-world example in which using simpler, less accurate, explanatory models results in improved prediction accuracy.

## 2.2   Illustrative Example

In this subsection we illustrate the way the bias-variance trade-off can impact the performance of predictive models with a familiar real-world example of "predicting the present" – the Spam filtering problem. A common approach to the task of distinguishing legitimate and Spam email messages is to construct a prediction model of the form

$$y = f_{class} \circ g_F(\mathbf{x}), \quad (3)$$

where each email message is encoded as a "bag of words" feature vector $\mathbf{x} \in \Re^{|V|}$, the entries of $\mathbf{x}$ are the (normalized) frequencies with which the words in the vocabulary V appear in the message, $g_F(.)$ is some model of the email, and $f_{class}(.)$ is the classifier, returning prediction $y = -1$ for legitimate email and $y = +1$ for Spam, respectively (Colbaugh and Glass 2012b).

It is assumed that n examples of legitimate and Spam emails $\{\mathbf{x}_i, y_i\}_{i=1}^n$ are available to use in building the predictor $f_{class} \circ g_F(\mathbf{x})$ (where $\mathbf{x}_i$ are emails and $y_i$ are the associated labels). We begin by deriving model $\mathbf{h} = g_F(\mathbf{x})$. Let $X \in \Re^{n \times |V|}$ denote the matrix obtained by stacking the emails $\mathbf{x}_i$ as rows, and factor X using the singular value decomposition (SVD):

$$X = \sigma_1 u_1 v_1^T + \sigma_2 u_2 v_2^T + \ldots + \sigma_r u_r v_r^T$$
$$\approx \sigma_1 u_1 v_1^T + \sigma_2 u_2 v_2^T + \ldots + \sigma_F u_F v_F^T, \quad (4)$$

where $\sigma_i$, $u_i$, $v_i$ are the singular values and (left and right) singular vectors of X, r is the rank of matrix X, $F < r$ is an integer, and (4) gives the optimal rank-F approximation of X (Hastie, Tibshirani, and Friedman 2009). The truncated SVD (4) of the matrix X of "training" instances allows a simple specification for the email model $g_F(\mathbf{x})$, namely $g_F(\mathbf{x}) = \mathbf{h} = [v_1^T \mathbf{x}, v_2^T \mathbf{x}, \ldots, v_F^T \mathbf{x}]^T \in \Re^F$. One advantage of this model for our purposes is that F provides a natural parameterization for the explanatory power of the model, with increasing F resulting in increased email model accuracy (see Figure 1).

There are a number of ways to learn the classifier $f_{class}(.)$ from the transformed training data $\{\mathbf{h}_i, y_i\}_{i=1}^n$, and we adopt a simple regularized least squares (RLS) approach (Hastie, Tibshirani, and Friedman 2009):

$$f_{class}(\mathbf{h}) = \text{sign}(w^T \mathbf{h}),$$

where $\mathbf{h}$ models the message of interest and $w \in \Re^F$ is the solution to the set of linear equations

$$[H^T H + \gamma I_F] w = H^T y,$$

and where matrix $H \in \Re^{n \times F}$ has email model vectors $\mathbf{h}_i$ for rows, $\mathbf{y} \in \Re^n$ is the corresponding vector of email labels, $I_F$ denotes the F×F identity matrix, and $\gamma \geq 0$ is a constant.

This setup enables straightforward investigation of the bias-variance trade-off quantified in (2). As F is increased in predictive model (3), it is expected that the explanatory accuracy of the model will increase (because more terms are retained in (4)), and it is of interest to examine the impact of this increased fidelity on prediction accuracy. In order to conduct such a study, we obtained a collection of 3000 legitimate

emails from various publicly-available sources, and added to this corpus a set of 3000 Spam emails acquired from B. Guenter's Spam trap (Colbaugh and Glass 2012b). The resulting dataset consists of 6000 emails composed of more than 250,000 words. By removing words which occur less than five times in the entire corpus, the vocabulary V is reduced to approximately 10,000 words.

The results of the study are displayed in Figure 1. The plot at the left of the figure shows that increasing the number of terms F in the SVD expansion (4) increases the explanatory accuracy of the model, and thereby decreases the bias term in (2), as expected. (The measure of explanatory accuracy used here is the normalized Frobenius norm of the matrix difference $X - (\sigma_1 u_1 v_1^T + \sigma_2 u_2 v_2^T + \ldots + \sigma_F u_F v_F^T)$). However, it can be seen from the plot at the right of Figure 1 that this increased model fidelity *does not* translate to increased predictive accuracy. Specifically, the plot shows how Spam/non-Spam prediction accuracy obtained with classifier (3) varies with parameter F (two-fold cross-validation on the Spam/non-Spam dataset described above). This plot reveals that the maximum accuracy is achieved with $F \approx 25$, and that prediction accuracy declines monotonically for larger values of F. Thus, although increasing F yields a better explanatory model, with smaller bias, this improved explanatory power comes at the cost of higher model complexity and an associated increase in estimation variance, and the latter results in reduced overall predictive power.

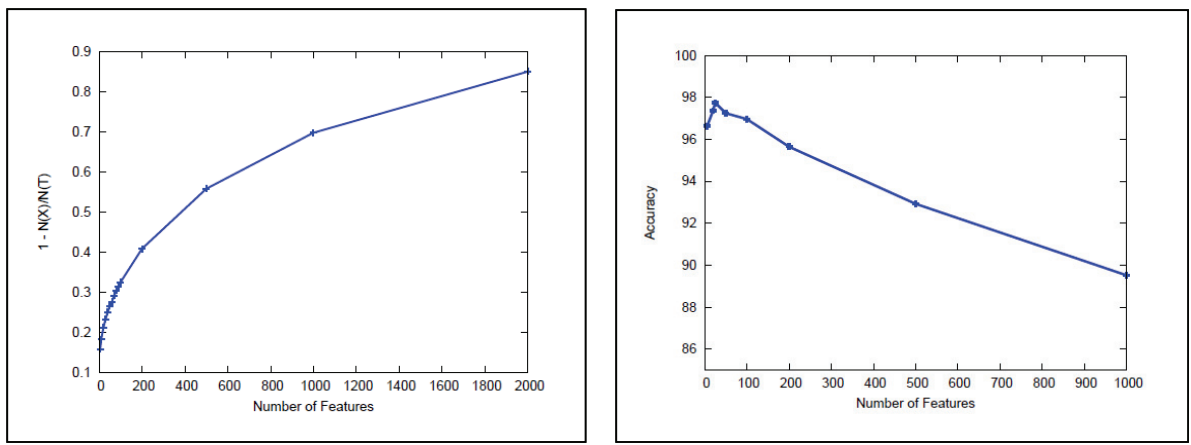

**Figure 1:** Spam/non-Spam modeling and prediction. The left plot shows that the accuracy of SVD-based email models (vertical axis) increases with the number of terms included in the SVD expansion (horizontal axis). The right plot demonstrates that this increased model fidelity does not guarantee better predictions: the accuracy of classifying emails as Spam or non-Spam (vertical axis) decreases with the number of terms retained in the SVD expansion (horizontal axis).

## 3    CASE STUDY: PREDICTING ADVERSARIAL INTERACTIONS

We have argued that predictive analysis is important for both the scientific development and practical applications of NESS, and have shown that good predictive models have attributes which distinguish them from the explanatory models that dominate the social sciences. In this section, we demonstrate through a case study that useful predictive analysis is *possible* to accomplish in real world settings. More precisely, this case study investigates signed social networks, where positive/negative edges reflect friendly/antagonistic social ties, and derives novel algorithms for two prediction tasks: 1.) predicting the signs of certain edges of interest (an example of "predicting the present"), and 2.) predicting the way adversarial networks will fracture under stress (an instance of "predicting the future"). Additional examples illustrating the feasibility of forming useful predictions for practically-important problems are given in (Colbaugh and Glass 2012a-c)

### 3.1 Problem Formulation

Social networks may contain both positive and negative relationships – people form ties of friendship and support but also of animosity or disapproval. These two types of social ties can be modeled by placing signs on the links or edges of the social network, with +1 and −1 reflecting friendly and antagonistic relationships, respectively. We wish to study the problem of predicting the signs of certain edges of interest by observing the signs and connectivity patterns of the neighboring edges. More specifically, for a directed social network $G_s = (V, E)$ with signed edges, where V and E are the vertex and edge sets, respectively, we consider the following edge-sign prediction problem: given some edge of interest $(u,v) \in E$ for which the edge-sign is "hidden", infer the sign of (u,v) using information contained in the remainder of the network.

It is natural to suspect that *structural balance theory* (SBT) may be useful for edge-sign prediction. Briefly, SBT posits that if $w \in V$ forms a triad with edge (u,v), then the sign of (u,v) should be such that the resulting signed triad possessing an odd number of positive edges; this encodes the common principle that "the friend of my friend is my friend", "the friend of my enemy is my enemy", and so on (Heider 1946). Thus SBT suggests that knowledge of the signs of the edges connecting (u,v) to its neighbors may be useful in predicting the sign of (u,v).

### 3.2 Prediction Algorithm

We approach the task of predicting the sign of a given edge (u, v) in the social network $G_s$ as a machine learning classification problem. The first step is to define, for a given edge, a collection of features which may be predictive of the sign of that edge. To allow a comparison with the (gold-standard) prediction method given in (Leskovec, Huttenlocher, and Kleinberg 2010), we adopt the same two sets of features used in that study. For a given edge (u,v), the first set of features defined in (Leskovec, Huttenlocher, and Kleinberg 2010) characterize the various triads to which (u,v) belongs. Because triads are directed and signed, there are sixteen distinct types (e.g., the triad composed of positive edge (u,w) and negative edge (w,v), together with (u,v), is one type). Thus the first sixteen features for edge (u,v) are the counts of each of the various triad types to which (u,v) belongs. Including these features is directly motivated by SBT. For example, if (u,v) belongs to many triads with one positive and one negative edge, it may be likely that the sign of (u,v) is negative, since then these triads would possess an odd number of positive edges and therefore be "balanced".

The second set of features defined in (Leskovec, Huttenlocher, and Kleinberg 2010) measure characteristics of the degrees of the endpoint vertices u and v of the given edge (u,v). There are five of these features, quantifying the positive and negative out-degrees of u, the positive and negative in-degrees of v, and the total number of neighbors u and v have in common (interpreted in an undirected sense). Combining these five measures with the sixteen triad-related features results in a feature vector $x \in \Re^{21}$ for each edge of interest (see (Leskovec, Huttenlocher, and Kleinberg 2010) for a more thorough discussion of these features and the motivation for selecting them). The feature vector x associated with an edge (u,v) will form the basis for predicting the sign of that edge.

We wish to learn a vector $c \in \Re^{21}$ such that the classifier orient $= \text{sign}(c^T x)$ accurately estimates the sign of the edge whose features are encoded in vector x. Vector c is learned, in part, from labeled examples of positive and negative edges. Additionally, the proposed learning algorithm leverages the insights of SBT. A simple way to incorporate SBT is to assemble sets $F^+$ and $F^-$ of positive and negative features, that is, sets of features which according to SBT ought to be associated with positive and negative edges, respectively. The triads to which (u,v) belongs in which the other two edges are positive are predicted by SBT to "contribute" to (u,v) being positive; thus the four features corresponding to triads with two positive labeled edges are candidates for membership in $F^+$ (there are four such features because $G_s$ is directed). Analogously, SBT posits that the eight features indexing triads in which exactly one of the two edges that neighbor (u,v) is positive are candidates for membership in $F^-$. (Note that the remaining four triad features index triads in which both of the edges neighboring (u,v) are negative, and as there is less empirical support for SBT in this case (Leskovec, Huttenlocher, and Kleinberg 2010) these features are not assigned to either $F^+$ or $F^-$.)

We now derive a machine learning algorithm for edge-sign prediction which is capable of leveraging SBT in its learning process. The development begins by modeling the problem data as a bipartite graph $G_b$ of edge-sign instances and features (see Figure 2). If there are n edges and 21 features, it can be seen that the adjacency matrix A for graph $G_b$ is given by

$$A = \begin{bmatrix} 0 & X \\ X^T & 0 \end{bmatrix} \quad (5)$$

where matrix $X \in \Re^{n \times 21}$ is constructed by stacking the feature vectors $x_i$ as rows, and each '0' is a matrix of zeros.

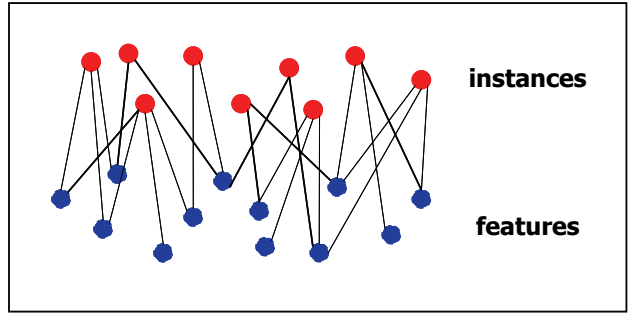

**Figure 2:** Cartoon of bipartite graph data model $G_b$, in which edge-instances (red vertices) are connected to the features (blue vertices) they contain, and link weights (black edges) reflect the magnitudes taken by the features in the associated instances.

Assume the initial problem data consists of a set of n edges, of which $n_l \leq n$ are labeled, and a set of labeled features $F_l = F^+ \cup F^-$, and suppose this label information is encoded as vectors $d \in \Re^{n_l}$ and $w \in \Re^{|F_l|}$, respectively. Let $d_{est} \in \Re^n$ be the vector of estimated signs for the edges in the dataset, and define the "augmented" classifier $c_{aug} = [d_{est}^T \; c^T]^T \in \Re^{n+21}$ that estimates the polarity of both edges and features. Note that the quantity $c_{aug}$ is introduced for notational convenience and is not directly employed for classification. More specifically, in the proposed methodology we learn $c_{aug}$, and therefore c, by solving an optimization problem involving the labeled and unlabeled training data, and then use c to estimate the sign of any new edge of interest with the simple classifier orient=sign($c^T x$). Assume for ease of notation that the edges and features are indexed so that the first $n_l$ elements of $d_{est}$ and $|F_l|$ elements of c correspond to the labeled data.

We wish to learn an augmented classifier $c_{aug}$ with the following three properties: 1.) if an edge is labeled, then the corresponding entry of $d_{est}$ should be close to this ±1 label; 2.) if a feature is in the set $F_l = F^+ \cup F^-$, then the corresponding entry of c should be close to this ±1 polarity; and 3.) if there is an edge $X_{ij}$ of $G_b$ that connects an edge x and a feature f and $X_{ij}$ possesses significant weight, then the estimated polarities of x and f should be similar. These objectives are encoded in the following optimization problem:

$$\min_{c_{aug}} \; c_{aug}^T L c_{aug} + \beta_1 \sum_{i=1}^{n_l} (d_{est,i} - d_i)^2 + \beta_2 \sum_{i=1}^{|V_l|} (c_i - w_i)^2 \quad (6)$$

where $L = D - A$ is the graph Laplacian matrix for $G_b$, with D the diagonal degree matrix for A (i.e., $D_{ii} = \Sigma_j A_{ij}$), and $\beta_1$, $\beta_2$ are nonnegative constants. Minimizing (6) enforces the three properties we seek for $c_{aug}$, with the second and third terms penalizing "errors" in the first two properties. To see that the first term enforces the third property, observe that this expression is a sum of components of the form $X_{ij}(d_{est,i} - c_j)^2$. The constants $\beta_1$, $\beta_2$ are used to balance the relative importance of the three properties. The $c_{aug}$ which minimizes objective function (6) can be obtained by solving the following set of linear equations:

$$\begin{bmatrix} L_{11} + \beta_1 I_{nl} & L_{12} & L_{13} & L_{14} \\ L_{21} & L_{22} & L_{23} & L_{24} \\ L_{31} & L_{32} & L_{33} + \beta_2 I_{|V_1|} & L_{34} \\ L_{41} & L_{42} & L_{43} & L_{44} \end{bmatrix} c_{aug} = \begin{bmatrix} \beta_1 d \\ 0 \\ \beta_2 w \\ 0 \end{bmatrix} \quad (7)$$

where the $L_{ij}$ are matrix blocks of L of appropriate dimension.

We summarize this discussion by sketching an algorithm for learning the proposed edge-sign prediction (ESP) classifier:

**Algorithm ESP**

1. Construct the set of equations (7).
2. Solve equations (7) for $c_{aug} = [\ d_{est}^T\ \ c^T\ ]^T$ (for instance using the Conjugate Gradient method).
3. Estimate the sign of any new edge x of interest as: orient = $sign(c^T x)$.

The utility of Algorithm ESP is now examined through an example involving edge-sign estimation for two social networks extracted from the Wikipedia online encyclopedia.

### 3.3  Wikipedia Example

This example examines the performance of Algorithm ESP for the problem of estimating the signs of the edges in two social networks extracted from Wikipedia (WP), a collectively-authored online encyclopedia with an active user community. We consider the following WP social networks: 1.) the graph of 103,747 edges corresponding to votes cast by WP users in elections for promoting individuals to the role of 'admin' (Leskovec, Huttenlocher, and Kleinberg 2010), and 2.) the graph of 740,397 edges characterizing editor interactions in WP (Maniu, Cautis, and Abdessalem 2011). In each network, the majority of the edges (≈80%) are positive. Thus we follow (Leskovec, Huttenlocher, and Kleinberg 2010) and create balanced datasets consisting of 20K positive and 20K negative edges for the "voting" network (Leskovec, Huttenlocher, and Kleinberg 2010), and 50K positive and 50K negative edges for the "interaction" network (Maniu, Cautis, and Abdessalem 2011).

This study compares the edge-sign prediction accuracy of Algorithm ESP with that of the impressive gold-standard logistic regression classifier given in (Leskovec, Huttenlocher, and Kleinberg 2010). The gold-standard algorithm is applied exactly as described in (Leskovec, Huttenlocher, and Kleinberg 2010). Algorithm ESP is implemented with parameter values $\beta_1 = 0.1$ and $\beta_2 = 0.5$, and with the vector w constructed using the four "positive triad" features $F^+$ and eight "negative triad" features $F^-$ noted above. As a focus of the investigation is evaluating the extent to which good prediction performance can be achieved even when only a limited number of labeled edges are available for training, we examine training sets which incorporate a range of numbers of labeled edges: $n_l$ = 0, 10, 20, 50, 100, 200.

Sample results from this study are depicted in Figures 2 and 3. Each data point in the plots represents the average of ten trials. In each trial, the edges are randomly split into equal-size training and testing sets, and a randomly selected subset of the training edges of size $n_l$ is "labeled" (i.e., the labels for these edges are made available to the learning algorithms). It can be seen that Algorithm ESP outperforms the gold-standard method on both datasets, and that the improved accuracy obtained with the proposed "SBT-informed" algorithm is particularly significantly when the number of labeled training instances is small.

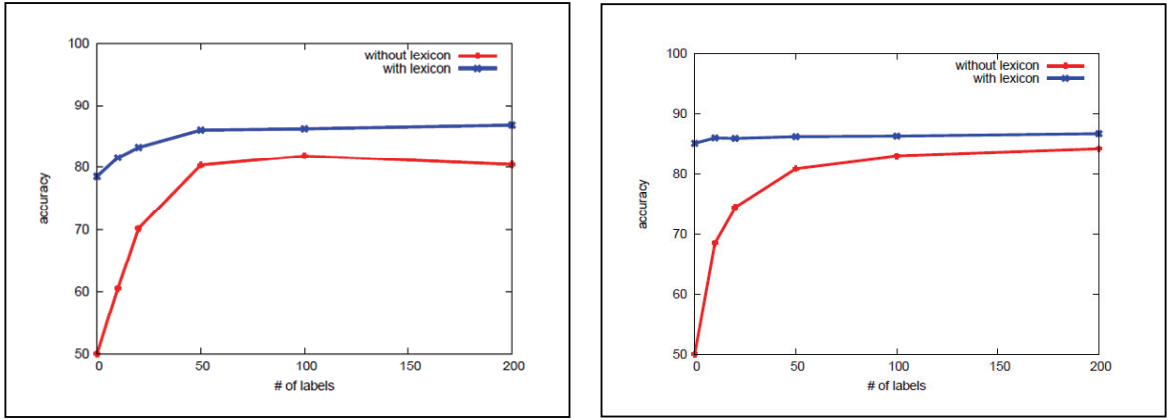

**Figure 3:** Results for WP "voting network" (left plot) and "interaction network" (right plot) studies. Each plot shows how edge-sign prediction accuracy (vertical axis) varies with the number of available labeled training instances (horizontal axis) for two classifiers: gold-standard (red) and Algorithm ESP (blue).

### 3.4  Network Fission Example

Recently it has been proposed that structural balance theory can be used to predict the way a network of entities (e.g., individuals, countries) will split if subjected to stress (Marvel et al. 2011), a capability of relevance in many applications. Briefly, (Marvel et al. 2011) models the polarity and intensity of relationships between the entities of interest as a completely connected network with weighted adjacency matrix $Z=Z^T\in\Re^{n\times n}$, where matrix element $z_{ij}$ represents the strength of the friendliness or unfriendliness between entities i and j. Note that this network model is somewhat more general than the one introduced above, in that each edge relating two individuals possesses both a sign and an intensity.

SBT is a "static" theory, positing what a stable configuration of edge-signs in a social network should look like. However, underlying the theory is a dynamical idea of how unbalanced network triads ought to resolve themselves to become balanced. A model which captures this underlying dynamics is given by the simple matrix differential equation (Marvel et al. 2011)

$$dZ/dt = Z^2, \quad Z(0)=Z_0. \tag{8}$$

To see the connection between these dynamics and SBT, observe that (8) specifies the following dynamics for entry $z_{ij}$:

$$dz_{ij}/dt = \sum_k z_{ik} z_{kj}.$$

Thus if triad {i,j,k} is such that $z_{ik}$ and $z_{kj}$ have the same sign, the participation of $z_{ij}$ in this triad will drive $z_{ij}$ in the positive direction, while if they have opposite signs then $z_{ij}$ will be driven in the negative direction. These dynamics therefore favor triads with an odd number of positive edge-signs, consistent with SBT (Heider 1946).

The paper (Marvel et al. 2011) proves that, for generic initial conditions $Z_0$, system (8) evolves to a balanced pattern of edge-signs in finite time; the balanced configuration is guaranteed to be composed of either all positive edges or two all-positive cliques connected entirely by negative edges. These configurations can be interpreted as predictions of the way a social network described by $Z_0$ will fracture if sub-jected to sufficient stress. More precisely, given a model $Z_0$ for a signed social network, model (8) can be used as the basis for the following two-step procedure for predicting the way the network will fracture: 1.) integrate (8) forward in time until it reaches singularity $Z_s$ (this singularity will be reached in finite time), and 2.) interpret $Z_s$ as defining a split of the network into two groups, where each group has all positive intra-group edges and the inter-group edges are all negative (and where one of the groups could be empty). See Figure 4 for an illustration of the dynamics of system (8).

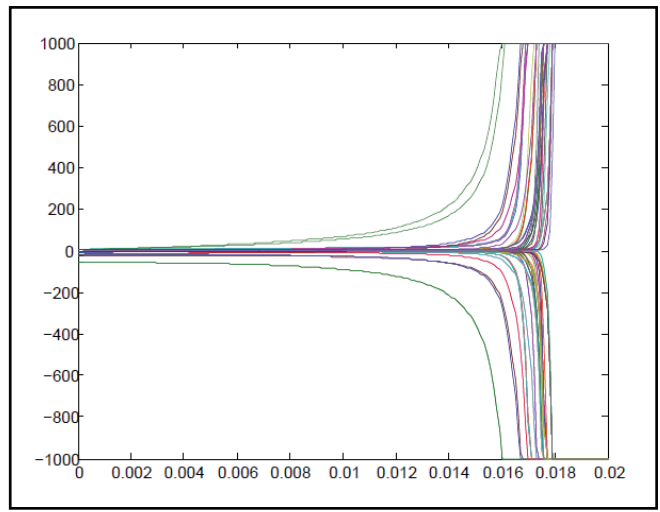

**Figure 4:** SBT dynamics. Simulation of model (8) initialized at the (scaled) "propensity" matrix given in (Axelrod and Bennett 1993) (horizontal axis is time and vertical axis is edge-weight).

Remarkably, (Marvel et al. 2011) shows that predictions obtained in this manner are in excellent agreement with two real-world cases of group fracture for which there is empirical data: the division of countries into Allied and Axis powers in World War II (Axelrod and Bennett 1993), and the split of the well-studied Zachary Karate Club into two smaller clubs (Zachary 1977). However, the analysis presented in (Marvel et al. 2011) requires that matrix $Z_0$ be completely known, that is, that all of the "initial" relationships $z_{ij}(0)$ between entities be measurable. Such comprehensive data are not always available in practical applications.

We have found that the requirement that relationship matrix $Z_0$ be perfectly known can be relaxed by using Algorithm ESP. More specifically, given a subset of the relationship data, the remaining weighted edge-signs can be predicted using Algorithm ESP, and these estimates $\underline{Z}_0$ can be used in place of $Z_0$ when initializing (8). We have tested this procedure using the relationship network proposed in (Axelrod and Bennett 1993) for 17 key countries involved in World War II. This investigation demonstrates that accurate prediction of which countries would eventually join the Allied forces and which would become Axis members can be made with less than 15% of the edge-signs known in advance. For example, data for only the relationships maintained by Germany and the USSR is sufficient to enable correct prediction of the ultimate alignment of all countries except Portugal.

## 4   CONCLUSIONS

This paper argues that predictive analysis is: 1.) an essential element of NESS, 2.) different from the more familiar task of explanatory modeling, and 3.) possible to achieve in important real-world applications. Future work will include predictive analysis of "complex contagion" events (Centola 2010, Colbaugh and Glass 2012a), involving the propagation of behaviors that are costly or controversial, and of various forms of adversarial dynamics.

## ACKNOWLEDGMENTS

This work was supported by the U.S. Department of Defense, The Boeing Company, and the Laboratory Directed Research and Development Program at Sandia National Laboratories. We thank Chip Willard of the U.S. Department of Defense and Anne Kao of Boeing for numerous helpful discussions on aspects of this research.

## AUTHOR BIOGRAPHIES

**RICH COLBAUGH** received a Ph.D. in Mechanical Engineering from The Pennsylvania State University in 1986, and is presently a Distinguished Member of the Technical Staff at Sandia National Laboratories. Rich's research activities focus on modeling, analyzing, predicting, and influencing the evolution of large-scale dynamical systems, with an emphasis on complex networks of relevance to advanced technology, social systems, and biology. His work has been recognized through awards from NASA, DoD, DOE, IEEE, SIAM, ASME, and AACC.

**KRISTIN GLASS** received a Ph.D. in Industrial Engineering from New Mexico State University in 1993, and is presently a Senior Research Scientist with ICASA at New Mexico Tech. Her research activities have focused on modeling, simulation, and analysis of complex social and technological systems. A recent emphasis of this work includes identifying and quantifying network properties which enable uncertainty reduction, reliability enhancement, and predictive analysis. Her work has been generously supported by NASA, DoD, DOE, ARO, and DARPA.

**CURTIS JOHNSON** leads the Analytics and Cryptography Department at Sandia National Laboratories. This group is focused on cybersecurity and social analysis. Curtis' research activities focus on the causes of human behavior and evolutionary, co-adaptive, and other complex adaptive systems.